# Curved beam generation and its experimental realization by rectangular prism with asymmetric polynomial back surface

Berkay Neşeli[1], Hamza Kurt[1] and Mirbek Turduev[2]

[1]School of Electrical Engineering, Korea Advanced Institute of Science and Technology (KAIST), Daejeon, 34141, Republic of Korea
[2]Department of Electrical and Electronics Engineering, Kyrgyz-Turkish Manas University, Bishkek 720038, Kyrgyzstan

**ABSTRACT**

With the discovery of self-accelerating beams, possibility of obtaining curved light beams in free space has been realized. These special beams paved the way for many new applications as well as the exploration of novel beam types. Recently, great research effort has been conducted to realize different types of curved beams such as photonic hook and airy beam. These curved types of beams are obtained by introducing structural asymmetry or applying non-uniform dielectric distribution to the input and output face of the structure. With this regard, we propose specially designed asymmetric structure with polynomial back surface which generates curved light beams. Proposed lossless dielectric structure can generate curved beams at frequencies varying from 15.78 GHz to 20.09 GHz and corresponding curvature angles of minimum 41.34° and maximum 57.58°, respectively. The physical background of the curved beam formation is based on interference of the exiting light waves that diffract on upper and bottom polynomial surfaces which provides phase modulation leading to the curved trajectory of the propagating light. In addition, the observed beam steering effect is further investigated and the experimental verification in microwave region is conducted to verify our design's operation principle.

## 1. INTRODUCTION

Investigating the behavior of propagating waves within and outside of the finite sized structures of special shapes is still one of the interesting research topics in the field of applied optics. It is well-known from the conventional optics that the propagation of electromagnetic waves inside the homogeneous medium follows the straight line along the propagation direction. Meanwhile, the study of Berry and Balazs in 1979 shows that the propagating wave can follow curved trajectory in free space which is named as Airy beam. They proposed that it is possible for a wave to accelerate without an exterior force only if the wave fits the Airy profile [1]. The discovery of new type of beams led to many important and remarkable achievement in generation of different types of beams [2]. What makes Airy beam so special and attractive are the features it displays such as their self-healing, accelerating and nondiffractive nature. Self-healing property of Airy beams allows them to reproduce their wave shape even if the beam faces any kind of disturbance or obstacles [3]. These features allow Airy beams to be implemented in optical communications [4], microscopy and imaging applications [5, 6]. The propagation of Airy beam in the curved trajectory is also another intriguing characteristic. In this regard, several studies related with self-accelerating curved beams have been reported [7-9]. One concern of these waves is that they are resulting from the solution to the paraxial wave equation, which restricts their curving performance such as limited pathlength and curving angles. For instance, at higher angles, these beams hardly maintain their optical shapes. However, this limitation has been overstepped by using a nonparaxial wave solution [10].

Properties discussed above have also pushed scientists to investigate the possibility of obtaining nondiffractive and curved beams with appropriate combination and modulation of optical devices. In this regard, other type of Airy-like beams has been explored; among them half Bessel [11], and Weber and Mathieu beams [12,13]. These Airy-like beams have wavefronts that follow ballistic trajectory, aka curved trajectory propagating with diffraction limited characteristics. On the other hand, the beams that follow ballistic trajectory can be generated by using microparticles. Here, microparticles act as micro-lenses which focus the incident light into a subwavelength volume which is known as 'photonic jet'

(PNJ) [14]. Also, further investigations show that PNJ can be obtained by using optical structures such as cuboids [15, 16] and even with plasmonic structures [17]. Strong localization and narrow beam properties of PNJ makes it possible to apply such beams to different areas such as microscopy applications [18], Raman spectroscopy [19], photolithography [20], and optical data storage [21].

Recently, alternative solution for obtaining curved beam named as "photonic hook" has been proposed [22]. Photonic hook is very attractive since it has the curved waveform and exhibits photonic jet's optical characteristics. The key difference between the newly proposed photonic hook and the ballistic trajectory beams is that the photonic hook is a curved beam with inflection point (Airy-like beams do not have such characteristic) which is obtained by combination of wedge prism and a cuboid structure [23, 24]. Moreover, Airy-like beams are principally generated in far-field but photonic hooks is a near field beam [22].

In this study, we numerically and experimentally verified the generation of the curved beam using compact, lossless, and all-dielectric structure having asymmetric polynomial profiled back surface. Performance of this structure is numerically analyzed and experimentally verified. Our design allows us to adjust the phase velocity and refracts the propagating light in such a way that resulting outgoing wave follows a curved trajectory in free space. Furthermore, beam steering effect is also observed and explained further in this study. These effects are investigated at selected frequencies of 22.1 GHz, 20.09 GHz, 18.41 GHz, 17.4 GHz, 15.78 GHz and 14.73 GHz. Experimental analysis is conducted in the microwave regime and obtained results confirming numerical findings are shared.

## 2. DESIGN APPROACH AND NUMERICAL RESULTS

The main objective of this study is to design and analyze a special optical device that allows light propagation in a curved trajectory. It is well known that to control the propagation behavior of the wave, its phase profile and energy flow should be manipulated accordingly. Here the phase of the incident wave can be modulated by refractive surfaces with the desired profile. For this reason, to understand the curved nature of such beams, we should analyze the background physical mechanism of the modulation of the incident wave phase profile provided by the refractive surface. For example, curved beam named as photonic hook, is realized by using a wedge-cuboid structure having asymmetry between input and output surfaces [25]. Here, a linear phase retardation is produced by introducing right triangular prism as a surface connected to a cuboid structure at the input face. Also, the prism is asymmetric with respect to the optical axis, and this results in the off-axis focusing with the curving effect. Moreover, one can obtain curved beam by symmetric geometry structure but with different dielectric compositions [26-28]. In other words, phase modulation is obtained by introducing index contrast in the symmetric structure. Regarding these, it is obvious that to force a wave to propagate in a curved trajectory one should modulate the surface or refractive index appropriately so that the structure induces asymmetric phase profile modulation with respect to optical axis. In the light of these observations, we came up with the idea of proposing a simple and effective approach to bend the light by using a prism having asymmetric polynomial back surface.

In this study, we want to note that we propose a novel conceptual structure for generation of a curved beam generating structure with a high curvature angle and not claim that we have generated another type of a photonic hook. If we make literature search, we can see that in studies related to photonic hooks or curved beams, authors mostly used refractive index difference of two materials [28], cubes attached to rectangular structures [29] or metal coating over cylindric structures [30], super-contrast dielectric microparticle incorporated with a flat mirror [31], microcylinder mounted with aluminum-mask [32], coherent illuminations [33], adjacent dielectric cylinders [34], or twin-ellipse mesoscale cylinders [35]. In our study, we present the new conceptual design of an asymmetric prism structure with polynomial surface for generation of curved beam by all-dielectric, lossless, and low refractive index medium without introducing any material change as well as external metallic coatings (without deteriorating homogeneity of the structure).

In Fig. 1(a), we present the schematic representation of the proposed structure with both size and surface asymmetry. As can be seen from the plot, back-surface profile $y_{surf}$ is formed by combination of two different polynomial functions (superimposed dashed lines). These polynomial functions are denoted as $y_{up}(x)$ and $y_{bot}(x)$ for the upper and bottom parts, respectively. Corresponding geometrical parameters of the polynomial surface prism (PSP) structure are given in a 3D model shown in Fig. 1(c). Here, thickness of the proposed structure is 6.80 cm, $x$ and $y$ spans of the structure are 3.50 cm and 13.60 cm, respectively. Also, the distance from the back surface to the location where upper and bottom polynomial surfaces meet at the optical axis is 4 cm. Since the experimental verification of the structure is intended to be realized in microwave region, the Polylactic Acid (PLA) material having refractive index of 1.60 is chosen.

Since we expect light deflection caused by prism back-surface, we applied ray analysis on our structure to see how polynomial surface effects the incident light as shown in Fig. 1(c). Here, velocities of deviating and converging rays are linked with the slopes of light rays leaving the structure. We were expecting rays to leave the structure to have negative and positive slopes corresponding to the exiting surfaces. Light rays leaving the proposed structure from the upper surface are refracted closely whereas rays leaving the bottom face refract apart from each other. Fig. 1(c) also tells us that all the rays leaving the structure meet each other at the location different from optical axis, i.e. within a region shifted from on-axis focusing location. It means that most of the optical energy is concentrated within that area and this caustic like region provides the desired light curving effect for this type of beams. Provided ray analysis show that asymmetric polynomial surface can be implemented for light curving purposes. Then as the next step, we have conducted the finite-difference time-domain (FDTD) [36] analysis of the proposed structure for a single frequency of 17.4 GHz to observe wave modulation effect. We used a plane wave source to excite the proposed structure having transverse magnetic (TM) polarization mode and the obtained electric field result is given in Fig. 1(d). This field plot indicates that the structure bends the light while keeping the majority of the optical energy inside the wave bending region.

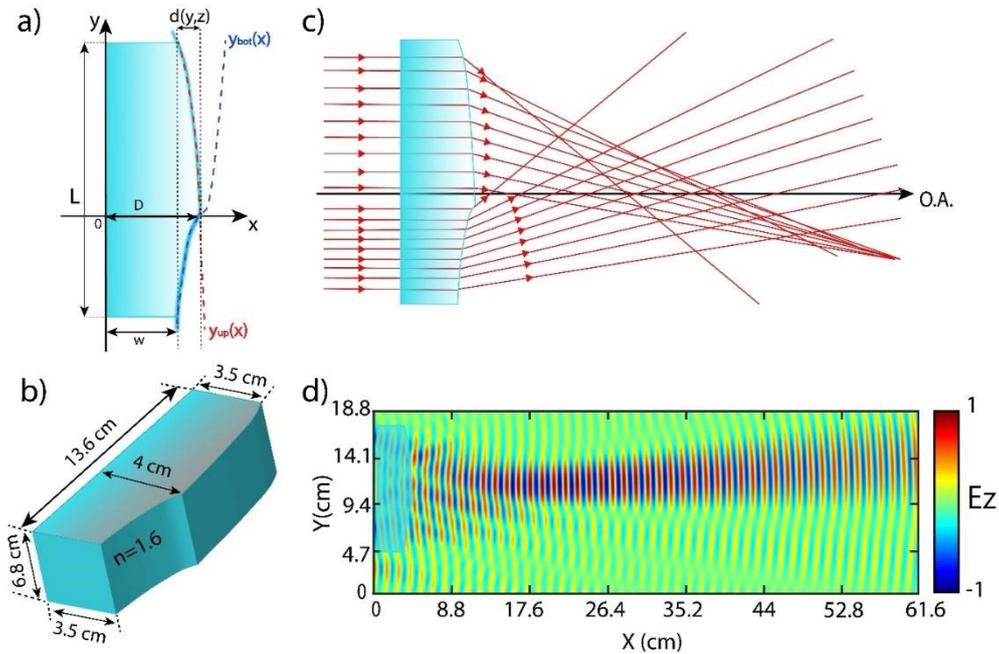

Figure 1. (a) Schematic representation of the proposed structure along with the profiles of the used polynomial functions. (b) 3D representation of the structure. (c) Ray tracing analysis of the PSP structure. (d) Field plot showing the electric field ($E_z$) profile of the resulting wave.

It is also important to analyze the effect of the polynomial surface on the phase transformation characteristic of the PSP structure. The surface polynomials are limited with the structural sizes of the PSP structure which is defined as follows:

$$y_{surf} = \begin{cases} y_{up}(x) = -a_3x^3 + a_2x^2 - a_1x + a_0, & D-w \leq x \leq D \\ y_{bot}(x) = b_3x^3 - b_2x^2 + b_1x - b_0, & D-w \leq x \leq D \end{cases}, (1)$$

where, $a_0$, $a_1$, $a_2$, $a_3$ and $b_0$, $b_1$, $b_2$, $b_3$ are the constant coefficients of the polynomials forming upper and bottom surface profiles. Also, D and $w$ are width parameters of the PSP as denoted in Fig. 1(a). In order to analytically represent the relation between polynomial surface and phase transformation effect the concept of complex transmittance can be used. The plane wave that propagates along the $x$-direction can be formulated as follows:

$$\Psi(x,y,z;t) = Ae^{-j(kx-\omega t)}. (2)$$

To show phase transformation of an incident plane wave interacting with the proposed PSP the complex transmittance function can be derived by using geometrical parameters of the PSP defined in Fig. 1(a) as:

$$t(y,z) = e^{-jkD}e^{-jkd(y,z)(n-1)} = h_0 e^{-jkd(y,z)(n-1)}, (3)$$

where $h_0$ is a constant phase factor. In this case the wave right after PSP structure will have the following wavefunction:

$$\Psi(D_+,y,z;t) = t(y,z) \times \Psi(0,y,z;t) = Ah_0 e^{-jkd(y,z)(n-1)}e^{j\omega t} = Ah_0 e^{-j(kd(y,z)(n-1)-\omega t)}. (4)$$

From Eq. 4, we can infer that the wavefronts after the PSP structure is directly depending on the surface function $y_{surf}$ of the structure since the variable width $d(y,z)$ of the PSP is also dependent on $y_{surf}$.

Next, we analyzed the PSP structure in a wider frequency region between 14.73 GHz and 22.1 GHz to observe wave curving characteristic with respect to the frequency change. The electric field intensity distributions for the selected frequencies are presented in Fig. 2. Superimposed black curves on the field distributions in Fig. 2 are indicating the variation of locations of the maximum intensities along the propagation direction. These lines are also another proof that optical energy follow the curved trajectory while its characteristic changes according to frequency. In the case of 22.1 GHz, despite the weak light curving performance, the beam is clearly lifted from the optical axis. On the other hand, curved profile is clearly observed for the frequencies 20.09 GHz, 18.41 GHz, 17.4 GHz and 15.78 GHz as can be seen from Figs. 2(b), 2(c), 2(d), and 2(e), respectively. In these frequency values, locations of the maximum intensities indicate the obvious curved profile and steering characteristic of the beam depending on operating frequency. It is important to note that incident wave is initially localized and then starts bending. Also, most of the energy is kept inside this curved path where light is localized and as expected, intensity decreases along the propagation direction. Moreover, side lobe intensities are weakly evident, and they increase as frequency decreases, but it still is negligible when compared to the intensity of the main lobe (curved part). At frequencies smaller than 15.78 GHz, the bending of the output beam starts to get distorted, and it comes closer to optical axis as can be inferred from Figs. 2(e) and 2(f).

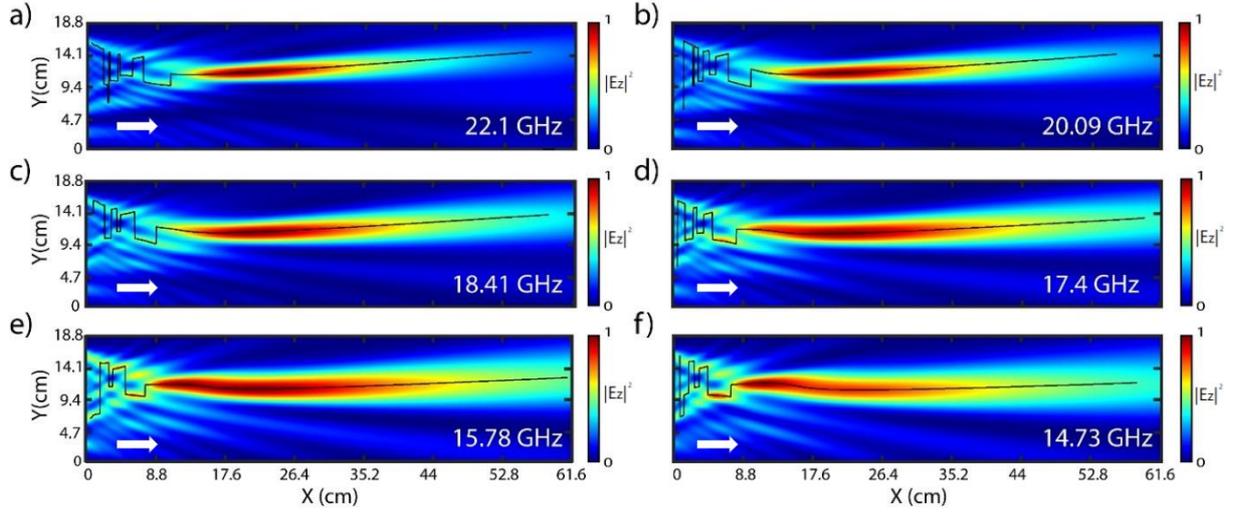

Figure 2. Obtained electric field intensity ($|E_z|^2$) distributions for frequencies 22.1 GHz, 20.09 GHz, 18.41 GHz, 17.4 GHz, 15.78 GHz and 14.73 GHz are given in (a), (b), (c), (d), (e) and (f), respectively. White arrows indicate the propagation direction of the incident light.

As it can be seen from the location of the maximum intensities, output beam follows curved trajectory and this characteristic can be analyzed by using simple curve angle calculation concept where degree of curvature can be considered. Fig. 3(a) is prepared to analyze the beam curving performance quantitatively where the 3D plot of electric field intensity at 17.4 GHz is presented. As schematically described in the corresponding plot, to measure the degree of curvature ($\theta$) we chose three points along the curve (along the maximum intensity points of the curved beam) as *A*, *B* and *C* but *A* and *C* points are chosen to have the same locations along the *y*-axis (to approximate circular curve). Next, the tangent lines going through the points *A* and *C* as well as the lines perpendicular to these tangents are defined. These perpendicular lines are intersecting each other at the point of *P*. Here, the angle between intersecting lines can be defined as degree of curvature $\theta$. By using simple geometry, the formula for calculating the degree of curvature $\theta$ can be derived as follows:

$$\theta = 2 \times arctan\left(\frac{AC}{\sqrt{4AP^2 - AC^2}}\right). \quad (5)$$

Numerically calculated degree of curvature $\theta$ of the curved beam is presented in Fig. 3(b) with respect to operating frequencies. As seen on the graph, curvature degree increases until it reaches 17.4 GHz-18.41 GHz range. These results agree with the field plots in Fig. 2 since we can observe the enhanced curvature and distortion of curvature of the output beam as frequency decreases. Since the PSP structure not only bends the incident wave but also confines/focuses the light inside the curving region, the off-axis light focusing performance of the proposed PSP structure is also analyzed. Hence, calculated full width at half maximum (FWHM) values are plotted in Fig. 3(b). FWHM values are increasing from 14.73 GHz to 15.78 GHz. If we look at the electric field intensity distributions given in Figs. 2(a)-2(f), we can observe that as frequency decreases, the off-axis focal point is approaching towards to the PSP structure. It means that at the frequency values lower than 17.18 GHz, the focusing point becomes closer to the region where refractions from the structure strongly interacts with the refracted wave, resulting in deterioration of focal point (dividing into partial separated focal points). This is the possible reason for decreasing of the FWHM values at the frequencies greater 17.38 GHz.

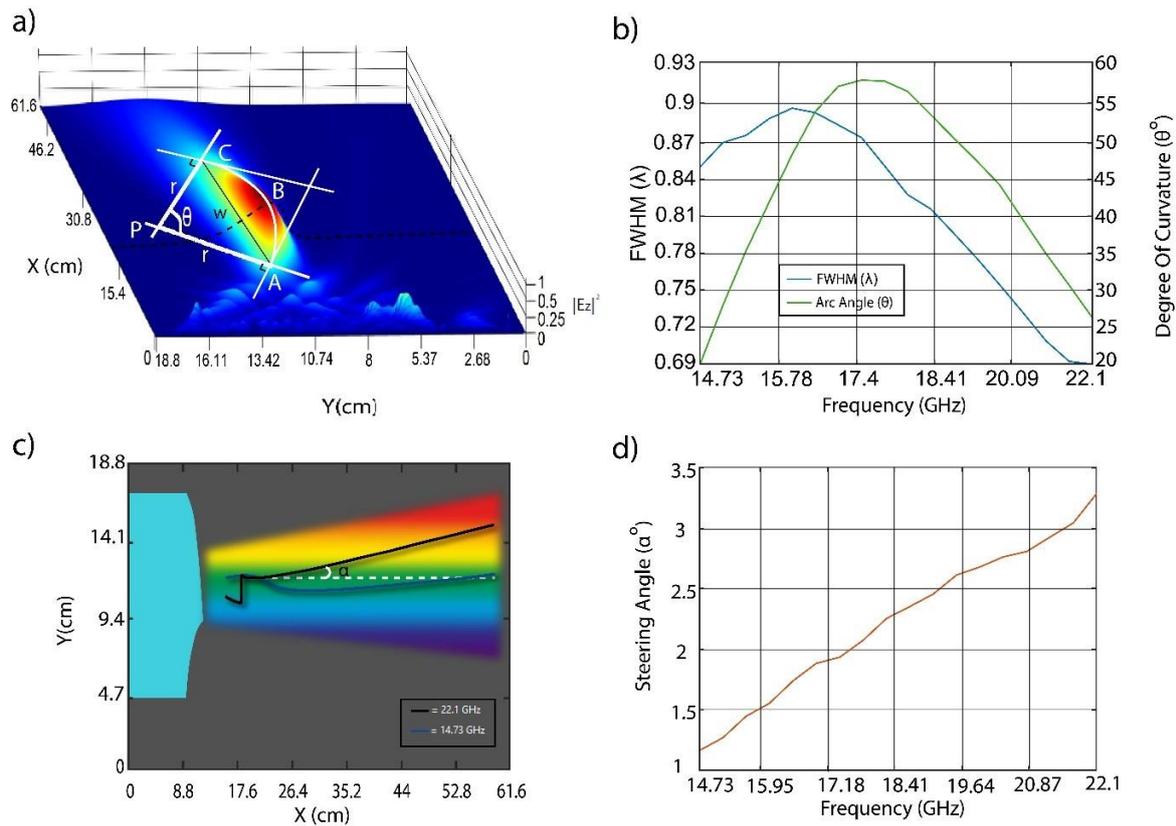

Figure 3. (a) 3D perspective plot of a resulting intensity plot. Cross-section at the maximum point is given as well as the curve angle calculation method. Representative letters (*A*, *B* and *C*) are also given to represent the chosen points for curve angle calculation. (b) Graphs of FWHM (λ) and degree of curvature (*θ*) are given with respect to frequency (GHz). (c) Depicts the beginning and ending of the peak intensity positions for different frequencies as well as the calculation method of steering angle (α) and (d) is the graph of resulting steering angle versus frequency.

Another interesting light manipulation behavior of PSP structure is its light steering ability with respect to operating frequency. Thanks to asymmetric polynomial surface we implemented, the incident wave is directed off axis and lifted from the optical axis as presented in Fig. 2. This effect can be applied for light steering applications in optics. As schematically described in Fig. 3(c), it can be observed that as frequency increases, the tilting of propagating wave increases. Steering angle *α* is defined as an angle of tilting of the beam with respect to optical axis and it varies from 1.2º to 3.3º for the frequencies between 14.73GHZ and 22.1GHz, respectively. Variation of the steering angle with respect to operating frequency is presented in Fig. 3(c). From the given result we can see almost linear dependency between *α* and frequency.

## 3. EXPERIMENTAL VERIFICATION OF THE PSP STRUCTURE

To experimentally verify numerical results of the curved beam generation by PSP structure, microwave region measurements are conducted. The designed PSP structure is fabricated by the 3D-FDM (fused deposition modelling) method with PLA polyester material. The refractive index of PLA material is $n=1.60$ at microwave region between 17 GHz and 18 GHz. Here, we prefer PLA as material for the structure because PLA is thermoplastic material, which is effectively lossless (possible attenuation caused by the material absorption can be neglected) in the microwave region. Also, PLA is cost efficient to fabricate the structure and perform microwave experiments. PSP structure is fabricated by using commercially available "MakerBot Replicator +" 3D printer by setting 100% infill ratio to generate the solid and homogeneous distribution of polyester throughout the structure. Figure 4(a) represents the schematic of the microwave experimental setup. Throughout the microwave experiment, Agilent E5071C ENA Vector Network Analyzer (VNA) is used to produce electromagnetic wave source through a horn antenna connected to VNA, whereas a monopole antenna is employed to scan the electric field distributions. In Fig. 4(a), schematic of the electric field distribution scanning area is presented as the Scanning Area for the *xy*-plane. The electric field distributions of the corresponding scanning area are swept by using motorized stage. Here, stage moves in *xy*- directions with the steps of 3 mm and 5 mm for *x*- and y- directions, respectively. The photographic illustration of the perspective and side views of the printed structure with corresponding dimensions is given in Fig. 4(b). Measured electric field intensity distributions at the back-plane of the PSP are given in Fig. 4(c) for frequencies of 17.2 GHz, 17.3 GHz, 17.4 GHz and 17.5 GHz. Also, the black arrow lines show the direction of energy flow of the output wave along the propagation *x*-direction. As it is noticed from the fields, wave is initially directed downwards and as it propagates it follows curved trajectory. The same approach for calculating the degree of curvature is used and corresponding angles are calculated as 25.3°, 36.7°, 46.42° and 34.29° for of 17.2 GHz, 17.3 GHz, 17.4 GHz and 17.5 GHz, respectively. Obtained results have matching points with the numerical results such as majority of the energy is kept within the curved profile and the degree of curvature increases until a certain frequency is reached and then decreases.

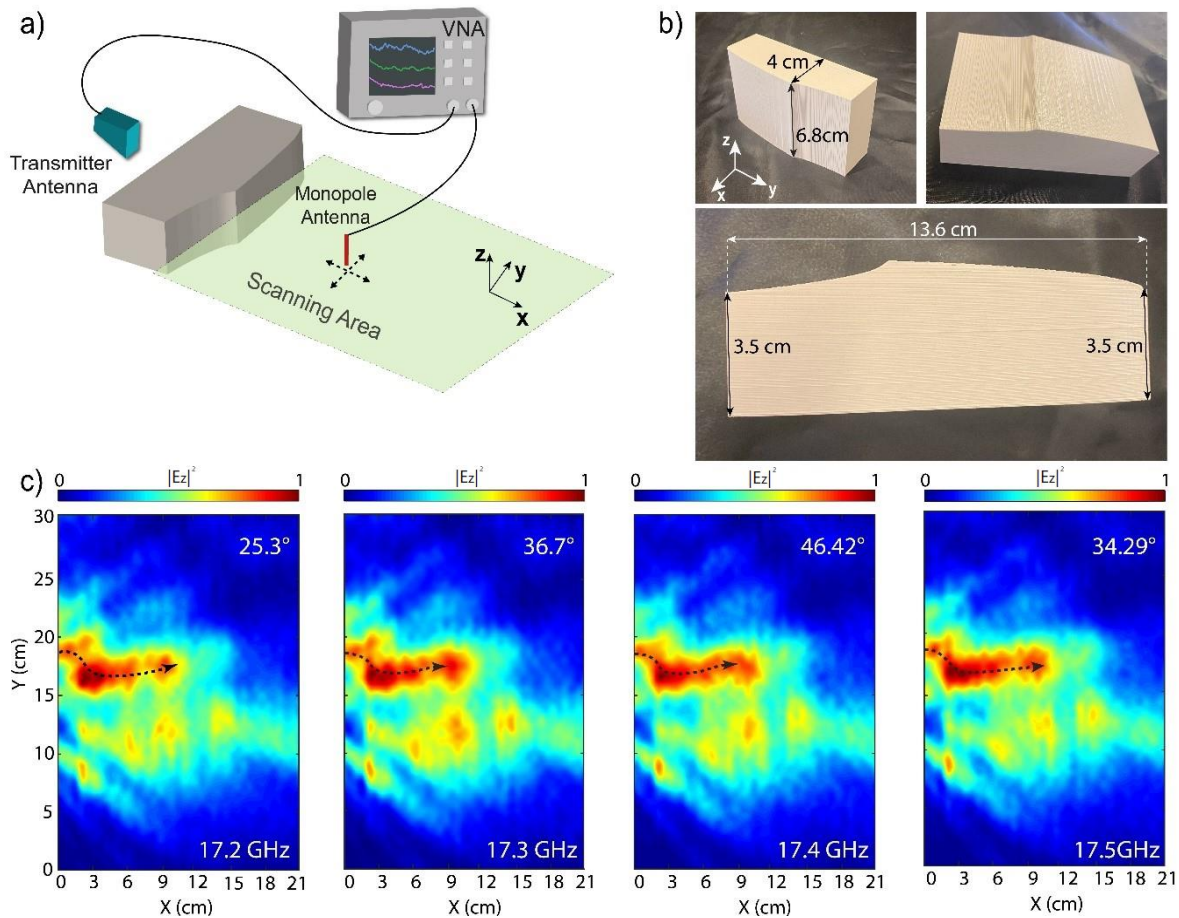

Figure 4. (a) Schematic representation of the experiment setup showing the scanning area. (b) Pictorial description of the printed structure designed to operate at 17.4 GHz is given with corresponding dimensions. (c) Obtained $|E^2|$ field plots for 17.2 GHz, 17.3 GHz, 17.4 GHz and 17.5 GHz are given as well as their calculated arc angle values.

There are few possible reasons why experimental results are not in complete match with numerical results. The first possible reason can be due to the possible manufacturing mismatch between numerically designed and fabricated structures. In Fig. 4(b), it is clear that the bottom of the printed structure is not completely flat but it is tilted which can be considered as a fabrication defect. Many printed structures have failed due to an imperfect bed adhesion problem and this structure is the best one among all of them. This inclined surface would affect the interaction of the incoming wave with our structure, which is not desired. Another possible reason is that our antenna was not able to produce a complete incident plane wave. Region between 17 GHz -18 GHz is measured as the range closest to a plane wave profile so that's why the experimental results have a narrow frequency range. To achieve plane wave profile, the antenna and the structure were placed 40 cm apart, which would result in energy loss of the incoming wave until it reached the designed structure. This is the possible reason why resulting waves are travelling shorter distances compared to numerical results. Despite these conditions, obtained results display a curved beam profile and verifies operating principle of proposed curved beam generating PSP structure.

## 4. CONCLUSION

The concept of "photonic hook" shows us that by applying various asymmetric arrangements on the structure, it is possible to manipulate the wave phase to achieve curved profile of the incident beam. In this study, we propose the new conceptual design of an asymmetric prism structure with polynomial surface for generation of near-field curved beam by all-dielectric, lossless, and low refractive index medium without introducing any material change as well as external metallic coatings (without deteriorating homogeneity of the structure) and its experimental realization for the first time. In addition, comparing to the curved beam generation concepts as discussed in the manuscript, our design concept is simple and practical. Also, proposed approach supported by analytical formulations and numerical simulations (both ray and wave based). Proposed structure operates at the frequency range between 14.73 GHz and 22.10 GHz. The degree of the curvature of the deflected beam is calculated according to the curve angle calculation concept and it varies between 41.34° and 57.58° within the frequency region of 15.78 GHz and 22.10 GHz. As frequency decreases from the first frequency value where bending is observed, the bending angle starts to decrease as well, and the curved profile of the beam gets distorted. Then by printing the proposed structure by using a 3D Printer with PLA material, we have conducted microwave experiments. Here, we observed similar effects, the energy is kept within the wave profile and the bending angle changed with frequency. For the frequencies of 17.2 GHz, 17.3 GHz, 17.4 GHz and 17.5 GHz the calculated curvature degrees are 25.3°, 36.7°, 46.42° and 34.29° respectively. We believe that our work can provide a new and practical way to generate curved beams. Also, generated curved beam has the potential in wide fields of applications, including optical trapping, subwavelength imaging, and signal switching. Furthermore, the effects of the curved optical energy flow can take role in in nanoscopy of biological cells, in nanoparticle trappings [37], in lab-on-a-chip microfluidics [38] and micro-channeling systems [39]. Last but not least, we believe that these results can give new ideas to researchers and pave way for new designs in the field of curved beams and photonic hooks.

## 5. REFERENCES

[1] Berry M V, & Balazs N L 1979 Nonspreading wave packets. *American Journal of Physics*, *47*(3).

[2] Ellenbogen T, Voloch-Bloch N, Ganany-Padowicz A, & Arie A 2009 Nonlinear generation and manipulation of Airy beams. *Nature Photonics*, *3*(7), 395–398.

[3] Zhang L, Ye F, Cao M, Wei D, Zhang P, Gao H & Li F 2015 Investigating the self-healing property of an optical airy beam. *Optics Letters*, *40*(21), 5066.

[4] Zhu G, Wen Y, Chen Y, Wu X, Liu J, Zhang Y, & Yu S 2017 Optical communications over obstacles by Applying two-dimensional ballistic-trajectory Airy beams. *Asia Communications and Photonics Conference*.

[5] Wang J, Hua X, Guo C, Liu W, & Jia S 2020 Airy-beam tomographic microscopy. *Optica*, *7*(7).

[6] Guo Y, Huang Y, Li J, Wang L, Yang Z, Liu J, Peng X, Yan W, & Qu J 2021 Deep penetration microscopic imaging with non-diffracting airy beams. *Membranes*, *11*(6), 391.

[7] Kaminer I, Segev M, & Christodoulides D N 2011 Self-accelerating self-trapped optical beams. *Physical Review Letters*, *106*(21).

[8] Dolev I, Kaminer I, Shapira A, Segev M, & Arie A. 2012 Experimental observation of self-accelerating beams in quadratic nonlinear media. *Physical Review Letters*, *108*(11).

[9] Bekenstein R, & Segev M. 2011 Self-accelerating optical beams in highly nonlocal nonlinear media. *Optics Express*, *19*(24), 23706.

[10] Kaminer I, Bekenstein R, Nemirovsky J, & Segev M. 2012 Nondiffracting accelerating wave packets Of MAXWELL'S Equations. *Physical Review Letters*, *108* (16), 163901.

[11] Li Y, Qiu C, Xu S, Ke M, Liu Z. 2015 Theoretical Study of Large-Angle Bending Transport of Microparticles by 2D Acoustic Half-Bessel Beams. *Scientific Reports* **5**, 13063.

[12] Aleahmad P, Miri M.-A, Mills M S, Kaminer I, Segev M, D Christodoulides N. 2012 Fully Vectorial Accelerating Diffraction-Free Helmholtz Beams *Physical Review Letters* **109**, 203902.

[13] Zhang P, Hu Y, Li T, Cannan D, Yin X, Morandotti R, Chen Z, Zhang X 2012 Nonparaxial Mathieu and Weber Accelerating Beams *Physical Review Letters* **109**, 193901.

[14] Luk'yanchuk S. B, Paniagua-Domínguez R, Minin I, Minin O, Wang Z. 2017 Refractive index less than two: photonic nanojets yesterday, today and tomorrow [Invited] *Optical Materials Express* **7**, 1820-1847.

[15] Liu C-Y 2015 Photonic jets produced by dielectric micro cuboids Appl. Opt. 54 8694–9.

[16] Pacheco-Peña V, Beruete M, Minin I. V., Minin O. V. 2014 Terajets produced by dielectric cuboids *Appl. Phys. Lett.* **105**, 084102.

[17] Minin I V, Minin O V, Glinskiy I A, Khabibullin R A, Malureanu R, Lavrinenko A V, Yakubovsky D I, Arsenin A V, Volkov V S and Ponomarev D S 2020 Plasmonic nanojet: an experimental demonstration Opt. Lett. 45 3244–7

[18] Hao X, Liu X, Kuang C, Li Y, Ku Y, Zhang H, Li H and Tong L 2013 Far-field super-resolution imaging using near-field illumination by micro-fiber Appl. Phys. Lett. 102 013104

[19] Huang S H, Jiang X, Peng B, Janisch C, Cocking A, Özdemir S¸ K, Liu Z and Yang L 2018 Surface-enhanced Raman scattering on dielectric microspheres with whispering gallery mode resonance Photon. Res. 6 346–56.

[20] Chang C-H, Tian L, Hesse W R, Gao H, Choi H J, Kim J-G, Siddiqui M and Barbastathis G 2011 From two-dimensional colloidal self-assembly to three-dimensional nanolithography Nano Lett. 11 2533–7

[21] Kong S-C, Sahakian A, Taflove A, & Backman V 2008 Photonic nanojet-enabled optical data storage. *Optics Express*, *16*(18), 13713.

[22] Geints Y E, Minin I V, & Minin O V 2020 Tailoring 'PHOTONIC hook' from janus dielectric microbar. *Journal of Optics*, *22*(6), 065606.

[23] Minin I V, Minin O V, Katyba G M, Chernomyrdin N V, Kurlov V N, Zaytsev K I, Yue L, Wang Z, & Christodoulides D N 2019 Experimental observation of a photonic hook. *Applied Physics Letters*, *114*(3), 031105.

[24] Gu G, Shao L, Song J, Qu J, Zheng K, Shen X, Peng Z, Hu J, Chen X, Chen M, & Wu Q 2019 Photonic hooks from Janus microcylinders. *Optics Express*, *27*(26), 37771.

[25] Minin I V, Minin O V, Ponomarev D S, & Glinskiy I A 2018 Photonic hook plasmons: A new curved surface wave. *Annalen Der Physik*, *530*(12), 1800359.

[26] Gu G, Zhang P, Chen S, Zhang Y, & Yang H 2021 Inflection point: A perspective on photonic nanojets. *Photonics Research*, *9*(7), 1157.

[27] Liu C-Y, Chung H-J, & E H-P 2020 Reflective photonic hook achieved by a dielectric-coated concave hemicylindrical mirror. *Journal of the Optical Society of America B*, *37*(9), 2528.

[28] Peng Z, Gu G, Shao L, & Shen X 2020 Easily tunable long photonic hook generated from Janus liquids-filled hollow microcylinder. *arXiv: Optics*.

[29] Yue L, Minin V. O, Wang Z, Monks N. J, Shalin S. A, Minin V. I 2018 Photonic hook: a new curved light beam. Optics Letters **43**, 771-774.

[30] Tang, F.; Shang, Q.; Yang, S.; Wang, T.; Melinte, S.; Zuo, C.; Ye, R. Generation of Photonic Hooks from Patchy Microcylinders. Photonics 2021, 8, 466.

[31] Geints, Y. E., Zemlyanov, A. A., Minin, I. V., & Minin, O. V. (2021). Specular-reflection photonic hook generation under oblique illumination of a super-contrast dielectric microparticle. Journal of Optics, 23(4), 045602.

[32] Minin, I. V., Minin, O. V., Liu, C. Y., Wei, H. D., Geints, Y. E., & Karabchevsky, A. (2020). Experimental demonstration of a tunable photonic hook by a partially illuminated dielectric microcylinder. Optics Letters, 45(17), 4899-4902.

[33] Zhou, S. (2020). Twin photonic hooks generated from two coherent illuminations of a micro-cylinder. Journal of Optics, 22(8), 085602.

[34] Zhou, S. (2020). Twin photonic hooks generated from two adjacent dielectric cylinders. Optical and Quantum Electronics, 52(9), 1-8.

[35] Shen, X.; Gu, G.; Shao, L.; Peng, Z.; Hu, J.; Bandyopadhyay, S.; Liu, Y.; Jiang, J.; Chen, M. Twin photonic hooks generated by twin-ellipse microcylinder. IEEE Photonics J. 2020, 12, 6500609.

[36] Lumerical Inc, "FDTD Solutions," http://www.lumerical.com/tcad-products/fdtd/.

[37] Park, E., Jin, S., Park, Y., Guo, S., Chang, H., & Jung, Y. M. (2022). Trapping analytes into dynamic hot spots using Tyramine-medicated crosslinking chemistry for designing versatile sensor. Journal of Colloid and Interface Science, 607, 782-790.

[38] Ahn, C., & Choi, J. W. (2007). Microfluidics and their applications to lab-on-a-chip. Springer handbook of nanotechnology, 523.

[39] Zhou, P., He, H., Ma, H., Wang, S., & Hu, S. (2022). A Review of Optical Imaging Technologies for Microfluidics. Micromachines, 13(2), 274.